\documentclass[conference]{IEEEtran}
\usepackage[utf8]{inputenc}
\usepackage{amssymb}
\usepackage{amsmath}
\usepackage{graphicx}
\usepackage{array}
\usepackage{ragged2e}
\usepackage{arydshln}
\usepackage{xcolor}
\usepackage{subcaption}

\usepackage{url}

\title{Metadata-based Malware Detection on Android using Machine Learning}
{\author
	{\IEEEauthorblockN{Alexander Hefter\IEEEauthorrefmark{1},
			Christoph Sendner\IEEEauthorrefmark{1}, and
			Alexandra Dmitrienko\IEEEauthorrefmark{1}}
		\IEEEauthorblockA{\IEEEauthorrefmark{1}University of W\"urzburg, Germany
		}
}}

\begin{document}

\maketitle

\begin{abstract}
In the digitized world, smartphones and their apps play an important role. To name just a few examples, some apps offer possibilities for entertainment, others for online banking, and others offer support for two-factor authentication. Therefore, with smartphones also, sensitive information is shared; thus, they are a desirable target for malware.\\
The following technical report gives an overview of how machine learning, especially neural networks, can be employed to detect malicious Android apps based on their metadata. Detection based on the metadata is necessary since not all of an app's information is readable from another app due to the security layout of Android. To do so, a comparable big dataset of metadata of apps has been collected for learning and evaluation in this work.\\
The first section, after the introduction, presents the related work, followed by the description of the sources of the dataset and the selection of the features used for machine learning, in this case, only the app permissions. Afterward, a free available dataset is used to find an efficient and effective neural network model for learning and evaluation. Here, the fully connected network type consisting of dense layers is chosen. Then this model is trained and evaluated on the new, more extensive dataset to obtain a representative result. It turns out that this model detects malware with an accuracy of 92.93\% based on an app's permissions.
\end{abstract}

\section{Introduction}
\label{seq:introduction}
For many people, their smartphone with its apps is a constant companion and supporter in their everyday life. Sometimes the apps only entertain for example when listening to music or playing games, but in other cases also sensitive information has to be shared such as in mobile banking apps or with respect to two-factor authentication. It is necessary and important that this information cannot fall into the wrong hands and, therefore, the installed apps have to be checked if they are malware or not. To do so, another app that can detect malware could be created. However, due to the structure of modern smartphone operating systems, such an app has only limited access to data of other apps on a commercially available smartphone. An example of this accessible data is the manifest where metadata like the required permissions and activities of an app are recorded. To make use of this information in the malware detection process machine learning algorithms can be applied.

This report describes the collection of metadata of benign and malicious Android apps and the subsequent usage of the permission features for the classification between malicious and benign apps by neural networks, as it has been done in a practical course in the area of secure software systems.

First of all, related works are discussed that apply machine learning techniques with static and also dynamic features of apps to distinguish between malicious and benign apps. This is followed by a section about the sources of the created dataset of around 2,8 million metadata of apps. At this point, a closer look is taken at the extracted permissions which serve as input features for the neural networks. For the subsequent section several neural network models such as fully connected, recurrent, and convolutional neural networks are created and described. They have been evaluated on another smaller and free available permission dataset to obtain a model structure that is both, efficient and effective. In the final evaluation step, the chosen model is then trained and evaluated with the new dataset. A look at further work concludes the report.
\section{Related Work}
\label{seq:related_work}
Several articles discuss the detection of Android malware on the basis of different features of the apps using machine learning or other classical classification methods. A compact overview is given in Table \ref{tab0:overviewlit}.\\

The APK Auditor \cite{APKAuditor} performs a static analysis of apps in the form of permission features. This is done with the help of a dataset of 1853 benign and 6909 malicious applications. The system consists of a database that contains information about applications and their analysis results, an Android client for the user to start the analysis of an app, and a server that connects the client with the database and manages the analysis process. Here, a permission malware score for an app that shall be analyzed is computed, and a threshold is used to classify between malicious and benign samples. APK Auditor achieves an accuracy of 88\% and a specificity of 0.925. 

In \cite{PAMD}, a similar method is presented where a permission score for the apps being analyzed is computed, too. However, instead of a threshold, this method classifies with a decision tree and reaches with only 60 samples an accuracy of 85\%.

Moreover, convolutional neural networks have been applied for malware detection by permission features \cite{Aksakalli}. The studies achieve an accuracy of 96.71\% with a part of the AndroTracker dataset, exactly 3,933 benign and 4,421 malware samples. The results have been compared to the ones of an SVM-based scheme (95.9\% accuracy), a fully connected neural network approach (96.1\% accuracy), a k-nearest-neighbor (95.8\% accuracy), and a Naive Bayes classification (91.2\% accuracy).

Furthermore, in \cite{AndMal2020}, another convolutional network has been used for malware detection. To do so,  for an app, a binary vector reflecting the availability of static features, such as permissions, activities, and similar, is created. In the following step, the authors transform the vectors into two-dimensional gray images. These images serve as input for the convolutional network. The network is trained and evaluated with a dataset of around 200,000 malware and 200,000 benign application data and achieves an accuracy of 93.36\%.

Vinayakumar et al. \cite{LSTMRNN} proposed a \textit{long short-term memory recurrent neural network} (LSTM-RNN) for malware detection with permission features. Here, bag-of-words embeddings are used to create the input vectors of the network. After the feature extraction step in the LSTM layers, the output is fed into dense layers and finally classified with an overall accuracy of 89.7\%. The underlying dataset, the Cyber Security Data Mining Competition (CDMC 2016), contains 61,730 APK files with 583 permissions.

With a comparable big dataset of 1,152,750 benign and 1,279,389 malicious samples SeqDroid \cite{SeqDroid} is trained and evaluated. SeqDroid is a machine learning model that combines convolutional layers, recurrent layers in the form of GRUs, and dense layers to a network that handles feature vectors consisting of the package name, the certificate owner of the APK, the requested permission, and the intended actions. As this is a binary classification, different thresholds have been tested, and the results are represented as a ROC curve. One of these test results has, for example, a true positive rate of 0.977 at a false positive rate of 0.01. With the given test set size of 264,130 benign and 222,298 malicious APKs, this yields an accuracy of around 98.4\%.

DroidDetector \cite{droiddetector} applies machine learning in the form of a deep belief network (DBN) with the help of a dataset consisting of static and dynamic features of 880 malicious and 880 benign apps. Static features are, in this case, permissions and API calls. The dynamic features such as cryptography operations and network and file input/output are monitored with the help of the sandbox application DroidBox. The DBN achieves an overall accuracy of 96.76\% by using all the features. Meanwhile, an evaluation restricted to static features yields an accuracy of 89.03\%. Moreover, other machine learning algorithms have been tested, such as Naive Bayes (83.86\% accuracy), Multi-layer perceptron (88.52\% accuracy), and SVM (92.84\% accuracy) with the complete feature set.

SAMADroid \cite{SAMADROID} is another system that makes use of machine-learning techniques for malware detection. It consists of 3 levels of app analysis: in the first level, in addition to static features, dynamic features are extracted, exactly system call tracing of apps that run on an Android device. This ensures a higher detection rate. The second level is the analysis of local and remote hosts, where the local host extracts the dynamic features as log files and forwards them to the server. The remote host extracts static features and uses them together with the logs for the analysis of the behavior of the application. In the third level, feature vectors from the analyzed features are created and serve as input to a machine-learning algorithm on the server. The tests make use of Drebin's dataset with 5,560 and 179 malware families and achieve an accuracy of 98.97\%. The article also refers to some more systems that include dynamic features and which end up with better results than with a purely static analysis. However, dynamic feature analysis causes much system overhead and, therefore, is not done for this report.\\

\begin{table*}
\centering
\caption{Overview of related work}
\label{tab0:overviewlit}
	\scalebox{1}{

\begin{tabular}{|c|c|c|c|c|}
\hline
Article & Accuracy & Features & Size of Dataset & Classification scheme\\
\hline
\hline
Talha et al. \cite{APKAuditor} & 88\% & permissions & 8,762 & permission score\\
\hline
Giang et al. \cite{PAMD} & 85\% & permissions & 60 & permission score\\
\hline
Karabey Aksakalli \cite{Aksakalli} & 96.71\% & permissions & 8,354 & CNN\\
& 95.9\% &  &  & SVM\\
& 96.1\% &  &  & neural network\\
& 95.8\% &  &  & k-Nearest Neighbor\\
& 91.2\% &  &  & Naive Bayes\\
\hline
Rahali et al. \cite{AndMal2020} & 93.36\% & permissions, activites, & 400,000 & CNN\\
& & receivers, providers & &\\
\hline
Vinayakumar et al. \cite{LSTMRNN} & 89.7\% & permissions & 61,730 & LSTM-RNN\\
\hline
Lee et al. \cite{SeqDroid} & 98.4\% & package name, intent & 2,432,139 & combination of RNN \\
& & actions, certificate owner,  & & (GRU) and CNN\\
& & permissions& &\\
\hline
Yuan et al. \cite{droiddetector} & 96.76\% & permissions, API calls, & 1,760 & DBN\\
& 83.86\% & dynamic features & & Naive Bayes\\
& 88.52\% & & & Multi-layer perceptron\\
& 92.84\% & & & SVM\\
\hline
Arshad et al. \cite{SAMADROID} & 98.97\% & static and dynamic & 5,560 & machine learning\\
\hline
\end{tabular}
}
\end{table*}

As it has been seen, most of the presented schemes in this section use datasets that are comparatively small. For the training and evaluation of machine learning algorithms, however, it is necessary to have a sufficient big dataset to keep the variance of the evaluation in accuracy and other metrics low and, therefore, obtain a general and representative result. Thus, the next section considers the question of data acquisition and dataset creation in the case of malware detection.
\section{Dataset}
\label{seq:dataset}
When applying machine learning algorithms to a classification problem, a lot of data is needed to train and evaluate a model. With respect to the detection of Android malware, only a few free available datasets that include metadata like permission information of malicious and benign apps exist. These datasets are rather small, and the malware samples are usually several years old. Thus, they are not a good basis for a representative result. Therefore, the first part of this section describes the sources of the dataset that has been created during the practical course. The second part is about the selection of the permission features which are used for training. 

\subsection{Sources of the Dataset}
\label{seq:sources_dataset}
Basically, all app stores, such as Google's Playstore \cite{Playstore}, can serve as a source for malicious and benign apps. To identify malware in between these unclassified apps, one has to analyze the apps either by hand or by antivirus programs.\\

In the case of the here created dataset VirusTotal \cite{VirusTotal}, which identifies malicious applications with over 70 antivirus scanners, is mainly employed to divide between malware and benign apps. Since it is also possible to download the metadata of an app when having a hash sum of it, VirusTotal is used as a source for the dataset, too. Furthermore, VirusTotal offended a dataset of around 30,000 malware apps in a Google Drive folder that has been used in the dataset.

The AndroZoo project \cite{Androzoo,AndrozooPaper} provides a huge set of over 14 million apps obtained from several sources, such as the Playstore \cite{Playstore}. These apps are indexed with their hash sums and already proofed by VirusTotal. An app that has been detected as malware by more than four antivirus scanners is supposed to be malware. If it is undetected, it is seen as a benign sample. Around 1.2 million malicious and 1.4 million benign samples of this collection have been added to the dataset either by obtaining the metadata of the apps from VirusTotal \cite{VirusTotal} or by downloading the apps and extracting the metadata with the help of Androguard \cite{Androguard,Androguard2}.

The CCCS-CIC-AndMal-2020 dataset \cite{AndMal2020Dataset,AndMal2020,AndMal20202} is a collaboration project between the Canadian Institute for Cybersecurity (CIC) and the Canadian Centre for Cyber Security (CCCS) and contains metadata of around 200,000 malware apps. Unfortunately, the columns which represent the permissions and other features are not named, and thus, this dataset cannot be used directly. However, the md5 hashes of these apps are also provided, so the metadata of the malware part of AndMal-2020 has been downloaded from VirusTotal \cite{VirusTotal} and integrated into the dataset.

Moreover, around 1,000 malware apps from the Contagio Mobile mini dump \cite{contagio,contagio2} have also been added to the dataset. As Contagio Mobile is a pure malware repository, it is not proofed with VirusTotal if these are malware apps.\\

In fact, the resulting dataset includes 1,447,566 benign and 1,397,986 malware samples, where 139,798 apps of each of both classes have been randomly selected as the test set, the same number of apps as the validation set, and the rest as the training set.

\subsection{Permission Selection}
\label{seq:permission_selection}
Although all metadata of the apps has been extracted and added to the dataset, in the further process of this technical report, only the permission features have been used for the malware detection algorithms. Therefore, all permissions that are requested by apps of the dataset have been collected together with their number of occurrences in benign and malware samples. In fact, 502,331 different permissions exist inside the dataset.\\

\begin{table*}
\centering
\caption{Excerpt of permissions that exist in the dataset with number of occurences}
\label{tab1:permissions}
	\scalebox{1}{
\begin{tabular}{|c||c|c|c|}
\hline
Permission & Benign & Malware & Total\\
\hline
\hline
android.permission.INTERNET & 1,420,018 & 1,392,491 & 2,812,509\\
com.android.launcher.permission.READ$\_$SETTINGS & 17,360 & 207,213 & 224,573\\
com.kiosgame.fruitblaster.permission.C2D$\_$MESSAGE & 0 & 1 & 1\\
android.permission.CHANGE$\_$WIFI$\_$STATE & 108,094 & 565,942 & 674,036\\
com.xgbuy.xg.permission.JPUSH$\_$MESSAGE & 0 & 2,666 & 2,666\\
com.htc.launcher.permission.READ$\_$SETTINGS & 126,134 & 54,464 & 180,598\\
android.permission.SET$\_$ACTIVITY$\_$WATCHER & 127 & 2,360 & 2,487\\
com.tencent.qqlauncher.permission.READ$\_$SETTINGS & 1,286 & 21,684 & 22,970\\
android.permission.USE$\_$BIOMETRIC & 12,389 & 232 & 12,621\\
dianxin.permission.ACCESS$\_$LAUNCHER$\_$DATA & 752 & 18,434 & 19,186\\
com.webcraftbd.flickr.permission.C2D$\_$MESSAGE & 3 & 19 & 22\\
\vdots & \vdots & \vdots & \vdots\\
\hline
\end{tabular}
}
\end{table*}

Table \ref{tab1:permissions} shows an excerpt of this permission list. It can be seen that there are permissions, such as the INTERNET permission, which are used by nearly all of the apps. The major part of the permissions are C2D$\_$MESSAGE of different companies, and most of them are requested by one app only. Due to memory consumption, it is not possible to use all permissions in a classification scheme. Therefore, permissions that occur in less than 26 apps are filtered out. Moreover, permissions that only occur in malware apps such as the JPUSH$\_$MESSAGE or respectively only in benign samples are also removed from the filtered permission list. Otherwise, a machine learning algorithm would classify all apps that request such permissions as malicious or, in the opposite case, as benign after training, even if there are apps which are not in the dataset and for which this is not true. Finally, 2,137 permissions that fulfill the conditions are identified.\\

In a further step, this list of filtered permissions is then used for the permission dataset creation. To do so, for each app, a vector consisting of zeros and ones with length 2,138 is generated, where the first element is the class label, zero for benign, one for malicious, and each of the other elements represent one of the filtered permissions which serve as input of a classification algorithm. Here, an element is zero if the related permission is not present in the app and one in the opposite case. The prepared permission dataset serves as input for the training and evaluation of a machine-learning model. Before this is done, in the next section, an effective and efficient neural network model is chosen with the help of a free available dataset.
\section{Neural Network Model Selection}
\label{seq:model_selection}
For obtaining good results in the detection of malware by a machine learning algorithm, it is necessary to have an efficient and effective model. As traditional classification methods such as decision trees or k-nearest neighbors are slow and resource intensive in prediction steps after training, only neural network methods are considered. On the one hand, the training of these methods requires more performance than the traditional methods, but this can be outsourced on a server. On the other hand, the predictions that are done on smartphones or similar are finished more quickly and require less performance.\\

Although it is possible to do this first evaluation and selection with a subset of the new dataset, which has been described in the section before, for better traceability, the tests of this section are based on the \textit{Android Permissions dataset (2019)} \cite{androidpermissiondataset}. This dataset contains 398 permissions of around 50,000 benign and 10,000 malware apps, where 1,500 data vectors of each of the two classes are randomly chosen for the validation set and the same number of vectors for the test set. With the rest of the data, the networks are trained. As the training set is not balanced, the malware part of this set is duplicated several times to obtain as many malware samples as benign ones. This procedure has no effect on the result of the networks since validation and test set are unchanged, but by doing this way, the networks can learn a wider variety of benign examples.\\

Several networks have been created with the Keras framework of TensorFlow \cite{tensorflow} for the tests: two \textit{recurrent neural networks (RNN)}, once with two LSTM layers and once with a GRU layer, a \textit{convolutional neural network (CNN)}, a \textit{fully connected neural network (NN)}, and a network that consists of a GRU layer in parallel to convolutional layers. These networks are now described:\\

The LSTM network consists of two LSTM layers, both with 100 units and a dropout rate of 25\%. On top of this network, there are three dense layers with $1024\times 2048\times 1024$ neurons and a dense output layer with one neuron. The dense layers are equipped with staggered dropout rates of $25\% \times 50\%\times 75\%$ to prevent overfitting. Additionally, $L_2$ regularization of $0.002$ and ReLU activation is used for all layers except for the output layer, which has to be activated by Sigmoid. Other activation functions have also been tested, but their results are worse compared to the ones with ReLUs.

The GRU network is similar, but it consists of only one GRU layer with 150 units and a dropout rate of 20\% instead of two LSTM layers. The output of the layer is fed into the same dense network as in the LSTM case, with the only difference being that all layers are regularized with a weight of $0.0004$.

The third network type, CNN, has three one-dimensional convolutional layers for feature extraction in connection with the well-known dense network described above. The convolutional layers $5\times 80\times 30$ filters with kernel sizes of $10\times 5 \times 3$. Moreover, each of the layers is equipped with zero padding and is followed by a max pooling layer. The three pooling sizes are set to $2\times 3\times 2$.

Furthermore, a network has been created which has parallel to the before described convolutional and pooling layers, a GRU layer with 80 units and a dropout rate of 25\%. The results of both networks serve as input of the fully connected dense layers as above.

The last network, the NN, only consists of the fully connected dense layers with $1024\times 2048\times 1024$ neurons and dropout. In contrast to before, this time, additionally to the dropout rate, the $L_2$ regularization of the layers is staggered with the weightings of $0.0005\times 0.001\times 0.002$.\\

\begin{figure*}
\begin{subfigure}{0.33\linewidth}
  \centering
  \includegraphics[width=1\linewidth]{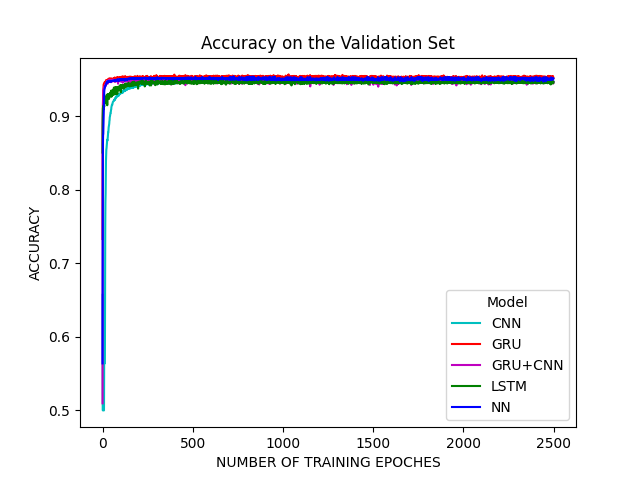}
  \caption{Complete curves}
\end{subfigure}%
\begin{subfigure}{0.33\linewidth}
  \centering
  \includegraphics[width=1\linewidth]{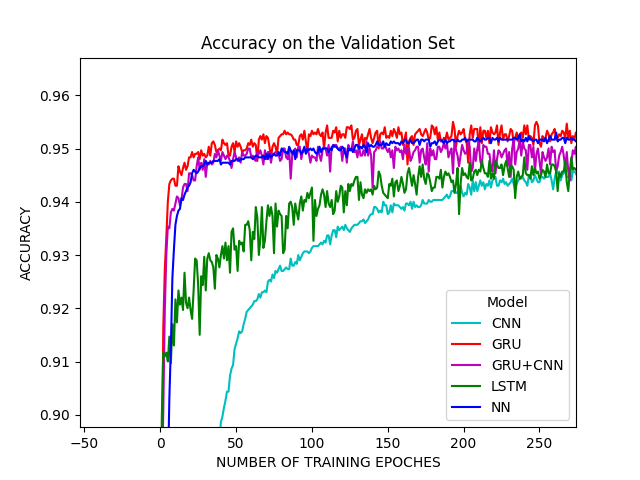}
  \caption{GRU converges fastest}
\end{subfigure}
\begin{subfigure}{0.33\linewidth}
  \centering
  \includegraphics[width=1\linewidth]{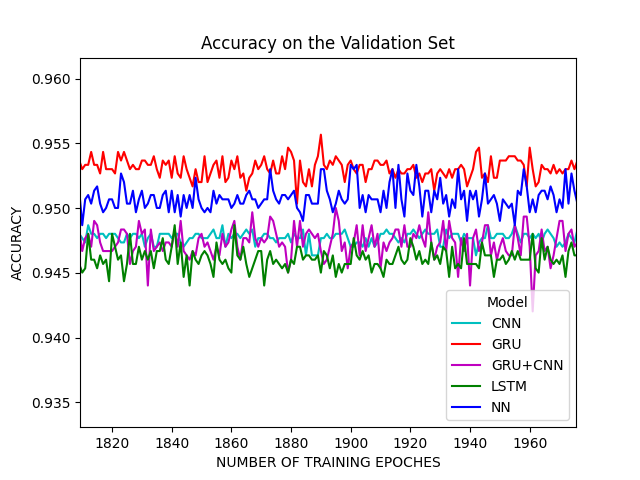}
  \caption{Saturation region}
\end{subfigure}
\caption{The change of accuracies during training}
\label{fig:AccuraciesOld}
\end{figure*}

In figure \ref{fig:AccuraciesOld}, the progress of accuracy at the validation set during the training process is shown. Evidently, the GRU network classifies the samples of this set best, followed by the fully connected network approach (NN). Moreover, the GRU network reaches the saturation region in fewer epochs than the other networks and, therefore, seems to have the highest convergence rate. However, as it can be seen in table \ref{tab2:resultsold}, the time consumption at one training epoch is for this network much higher than for the other networks, especially for the NN. Therefore, the NN converges much faster if one takes the total time consumption into account. Furthermore, it is surprising that LSTM, GRU+CNN, and CNN work worse than the NN since all these networks have included the NN as a part. Evidently, instead of additional feature extraction in the early layers, information gets lost. For networks that make use of convolutional layers, this can be explained by the following reason: As convolutions have only a limited range, namely the kernel size, their filters can only extract features that consist of neighboring data points. In this case, the data points are a bag of permissions, and in contrast to time series where data points have predecessors and successors, there exists no real ordering and also no autocovariance between the data points. This implies that if the arbitrary initial ordering of the permissions were changed before training, the overall result would be different since convolutional layers are not fully connected in contrast to recurrent or dense layers.\\

\begin{figure}
\centering
\begin{subfigure}{0.80\linewidth}
  \centering
  \includegraphics[width=1\linewidth]{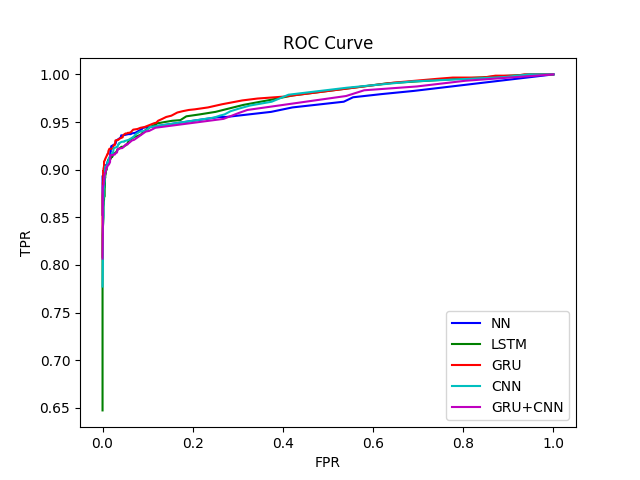}
  \caption{Complete ROC curve}
\end{subfigure}%
\\
\begin{subfigure}{0.80\linewidth}
  \centering
  \includegraphics[width=1\linewidth]{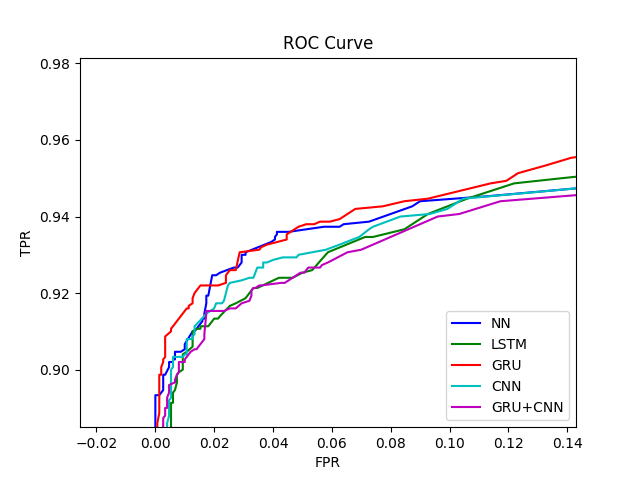}
  \caption{Enlargement on the best thresholds}
\end{subfigure}
\caption{ROC curves of the different networks types evaluated on validation set}
\label{fig:ROCOld}
\end{figure}

As the here-done classification consists of only two classes, malicious and benign, it is possible to draw ROC curves of the networks where every point is the result of another threshold applied to the output. In figure \ref{fig:ROCOld}, the ROC curves evaluated on the validation set are presented. Again the GRU network works best for most thresholds, but also, the NN is at the most significant positions, not much worse, but in some cases even better. The area under the curve (AUC) given in table \ref{tab2:resultsold} reflects this fact in numbers.\\

\begin{table}
\centering
\caption{Results of the test set with threshold 0.5 for all networks, AUC of the validation set, and epoch time during training}
\label{tab2:resultsold}
\scalebox{0.8}{
\begin{tabular}{|c||c|c|c|c||c|c|}
\hline
 & Accuracy & Recall & Precision & F1 & AUC & Epoch Time\\
\hline
\hline
NN & 0.9547 & 0.9220 & 0.9864 & 0.9531 & 0.9149 & $\sim$ 0.29 sec \\
\hline
GRU & 0.9533 & 0.9273 & 0.9782 & 0.9521 & 0.9216 & $\sim$ 18 sec \\
\hline
LSTM & 0.9500 & 0.9140 & 0.9849 & 0.9481 & 0.8971 & $\sim$ 16 sec \\
\hline
CNN & 0.9560 & 0.9220 & 0.9893 & 0.9545 & 0.9058 & $\sim$ 1,1 sec \\
\hline
GRU+CNN & 0.9547 & 0.9200 & 0.9885 & 0.9530 & 0.9032 & $\sim$ 13 sec \\
\hline
\end{tabular}
}
\end{table}

In table \ref{tab2:resultsold}, one can also find the evaluation results of the test set. For this evaluation, the standard threshold of 0.5 has been used. A threshold optimization would also be possible, but as the ROC curves of figure \ref{fig:ROCOld} show, this would not make a big difference. The GRU network has the highest recall and, thus, detects the most malware. However, the precision is not as high as for the other networks, and the CNN network has the highest accuracy with 95.60\%, followed by the NN with 95.47\%. Again the worst result is given by the LSTM network. It is also seen in the recall of the results that all networks are not able to identify a high percentage of malware of about 7-9\%, but nearly all identification that is done are correct, as the precision values show.\\

The aim of the evaluation in this section is to find a network type for malware detection that is both efficient and effective. If absolute training time counts, this is the case for the NN approach. Although the GRU network results have been better in most cases, due to time consumption, it is out of the scope of the practical course to train this network on the big dataset of section \ref{seq:dataset}. Therefore, in the next section, the NN network is trained and evaluated on this set.
\section{Evaluation on the new Dataset}
\label{seq:evaluation}
In the section before, the fully connected neural network consisting of dense layers has been identified as the optimal choice for the evaluation with the new dataset of section \ref{seq:dataset}. Two different fully connected networks are evaluated on this dataset, the one of the section before and a bigger one with more neurons, exactly $4096\times 8192 \times 4096$. The second network is used because, in most cases, more neurons can catch more features, and a network with more neurons is, therefore, more accurate. The dropout rates are kept in the same way staggered as before for both networks at $0.25\times 0.5\times 0.75$. The networks have been trained until the most important metrics, accuracy, recall, and precision, do not significantly change for more than six hours.\\

\begin{figure}
\centering
\begin{subfigure}{0.80\linewidth}
  \centering
  \includegraphics[width=1\linewidth]{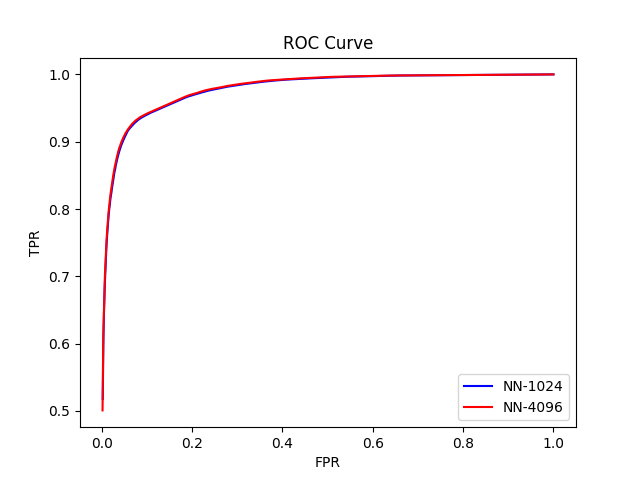}
  \caption{Complete ROC curve}
\end{subfigure}
\\
\begin{subfigure}{0.80\linewidth}
  \centering
  \includegraphics[width=1\linewidth]{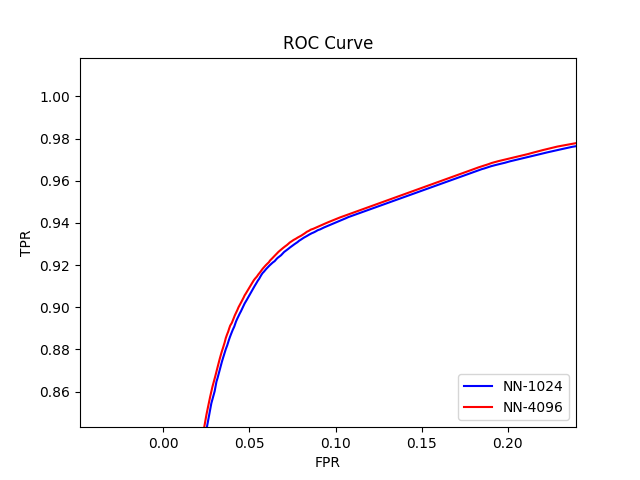}
  \caption{Enlargement on the best thresholds}
\end{subfigure}
\caption{ROC curves of the two fully connected networks evaluated on validation set}
\label{fig:ROCNew}
\end{figure}

\begin{table}
\centering
\caption{Results of the test set with threshold 0.5 for both networks, AUC of validation set, and epoch time during training}
\label{tab3:resultsnew}
\scalebox{0.8}{
\begin{tabular}{|c||c|c|c|c||c|c|}
\hline
 & Accuracy & Recall & Precision & F1 & AUC & Epoch Time\\
\hline
\hline
NN-1024 & 0.9285 & 0.9182 & 0.9375 & 0.9277 & 0.8887 & $\sim$ 9.5 sec \\
\hline
NN-4096 & 0.9293 & 0.9157 & 0.9413 & 0.9283 & 0.8929 & $\sim$ 60 sec \\
\hline
\end{tabular}
}
\end{table}

Figure \ref{fig:ROCNew} shows the ROC curves evaluated on the validation set. They are nearly identical. The results of the network with $4096\times 8192 \times 4096$ neurons are only a little bit better than the network that has only a quarter of the neurons in each layer. Moreover, the AUC of 0.8887 and 0.8929 differ not very much, but the time that one training epoch consumes: the bigger network needs about 60 seconds, the smaller one only 9.5 seconds.\\

In table \ref{tab3:resultsnew}, the results of both networks with respect to the test set are presented. Again the bigger network is slightly better at classification than the smaller one. Furthermore, it should be noticed that there is a big difference of over two percent in accuracies between the evaluation with this dataset and the evaluation with the dataset of section \ref{seq:model_selection}. Probably, the reason for this is the lower precision value which could be caused by apps that have the same permissions in the dataset but different labels. Moreover, the reduction of over 500,000 permissions to only 2,137 is, on the one hand, necessary, but on the other hand, the network has fewer possibilities to differentiate between the apps. One possibility to improve the detection rate is to add additional features, such as the certificate owner of an app. This feature is also available in the dataset of section \ref{seq:dataset} but has neither been used for learning nor for evaluation yet. By adding this feature, apps that are, for example, produced by a big company such as Google and that do not have a clear permission classification can be marked as benign. In the other case, apps of known malware producers can be marked as malware.
\section{Conclusion}
\label{seq:conclusion}
As it has been mentioned in the introduction smartphone apps are indispensable in many areas of life and malware producers exploit this fact for their own targets. The studies of the practical course this report belongs to show that it is possible to detect malware on the basis of requested permissions with high reliability. Several neural network classification schemes have been created and evaluated where the ones that are fully connected show the best results. A main point of the practical course has been the collection of a comparable huge dataset of benign and malware apps. The evaluation of the fully connected neural network with this dataset achieves an accuracy of 92.93 percent at a recall of 0.9157 and a precision of 0.9413. The work also shows that although the same classification scheme is applied results can deviate very much only by changing the dataset. In future works, schemes that also include the certificate owner of an app could be created to achieve a better classification result and, thus, a higher detection rate of malware. Moreover, as the tests on the smaller dataset show, recurrent networks especially with GRUs can be used as alternatives to dense networks, and, therefore, could be trained and evaluated with the here created dataset.

\bibliographystyle{plain}
\bibliography{references}

\end{document}